\documentstyle[12pt]{article}
\begin{document}

\begin{center}
{\Large \bf $d=5$ operators in SUSY GUT: \\
fermion masses versus proton decay \footnote{Talk presented by M.Vysotsky
at the Conference "Quarks 98", Suzdal, Russia, May 18-24, 1998.} }
\end{center}
\vspace{0.5cm}
\begin{center}
{\large Z.Berezhiani$^{a, b}$,
{}~Z.Tavartkiladze$^{a, b}$,
{}~M.Vysotsky$^{c, a}$}
\vspace{0.5cm}

$^a${\em Istituto Nazionale di Fisica Nucleare,
Sezione di Ferrara, 44100 Ferrara, Italy \\
$^b$ Institute of Physics, Georgian Academy of Sciences,
380077 Tbilisi, Georgia \\
$^c$ Institute of Theoretical and Experimental
Physics, 117259 Moscow, Russia}\\
\end{center}

\begin{abstract}
In the minimal SU(5) SUSY GUT $d =5$ operators lead to $p\to K^+
\nu$ decay with the proton life time of the order of $10^{28}$ years
for the natural choice of the parameters of the theory. This value
is in strong contradiction with experimental bound $\tau_{p\to K\nu}
> 10^{32}$ years. $d=5$ operators are induced by colored Higgsino
exchanges and are closely (through SU(5) and super symmetry)
related to another wrong prediction of SU(5) SUSY GUT: $m_d /m_s =
m_e /m_{\mu}$. We demonstrate how in the model where reasonable
pattern of quark and lepton masses and CKM mixing angles are
obtained proton decay can be suppressed and
proton life time can be
close to the present experimental bound.
\end{abstract}

\section{Introduction}
As everyone is aware of the most attractive candidate for the physics
beyond the Standard Model is low-energy supersymmetry
(see reviews \cite{1}). It
helps in solving hierarchy problem -- so, GUT's get firm theoretical
foundation. Experimental signal in favor of SUSY GUT comes from the
numerical value of electroweak mixing angle,
$\sin^2\theta^{exp}\approx 0.23$, which nicely coincides with SUSY GUT
value, while contradicts non-SUSY GUT prediction $\sin^2\theta
\approx 0.21$ \cite{2}. Another manifestation of this phenomena is
prediction of the $\alpha_s(M_Z) $ value which nicely coincides with LEP
and other low-energy measurements.

One of the most spectacular prediction of Grand Unification is proton
decay. In nonsupersymmetric theories proton decay through $d=6$
operators mostly via $p\to e^+\pi^0$ channel. Modern experimental
bound on this particular mode is $\tau_{p\to e^+\pi^0}\geq 10^{32}$
years which strongly contradicts prediction of the SU(5) GUT:
$\tau_{p\to e^+\pi^0}= 10^{28\pm 2}$ years \cite{3}. In
supersymmetric GUT's operators with $d=6$ are also generated. But
since in SUSY GUT's unification scale is approximately 30 times
larger than in non-SUSY GUT's, proton life time due to operators with
$d=6$ is of the order of $10^{34}$ years \cite{1}
which is beyond discovery possibilities because of background problems.

In supersymmetric models  the $d=4$ trilinear B and L violating
couplings can be introduced. They mediate  fast decay of proton.
That is why one should impose on the theory condition of absence of
such operators.

However, the $d=5$ operators
are induced in SUSY GUTs by exchange
of the color-triplet Higgsinos, which are partners of the
Higgs doublets in the GUT multiplets \cite{4}.

In the second section of this paper old
result \cite{5} for $\tau_{p\to K^+ \nu}$ in SU(5) SUSY GUT
will be reanalyzed.
Feynman diagrams which induce this decay are shown on Fig. 1.
This refreshment is necessary since at the time
when \cite{5} was written on the one hand lower experimental bound on
$\tau_{p\to K^+ \nu}$ was two orders of magnitude weaker than
now, on the other hand, neither $m_{top}$ was known nor
Kobayashi-Maskawa mixing angles were measured with modern accuracy.
Using these updated numbers together with old (but still valid) value
of decay matrix element from \cite{5} we get our central statement:
proton decays too fast and we cannot naturally be within experimental
bound \cite{6}:  $\tau_{p\to K^+\nu} > 10^{32}$ years.  In this way
we come to the following conclusion:  minimal supersymmetric SU(5)
GUT should be modified. In other words, some mechanism for
suppression of the nucleon decay amplitude is necessary and we are
not the only who shared this point of view \cite{7}.

In SO(10) theory suppression of the proton decay can occur for the
following two reasons:

1) The scalar sector is arranged in such a way, that the nucleon decay
parameter -- $(M_T^{-1})_{11}$ vanish or is strongly suppressed \cite {babu}.
SO(10) model in which
the proton decay is strongly suppressed and in which gauge
and fermion mass hierarchy was explained naturally was suggested in
\cite{9}.

2) Another possibility of  stabilizing the proton by implementing
the 45-plet with VEV towards $T_R$ direction in the Yukawa
sector was suggested in refs \cite {gia} \footnote {The analyses of
the nucleon decay in this model was presented in \cite{10}. }
and \cite {zura}.

Our approach here is different. We study $SU(5)$ SUSY GUT where
unsatisfactory relations of minimal theory $m_{\mu} =m_s$ and $m_e
=m_d$ at GUT scale are avoided.

These unsatisfactory
predictions come from the same Higgs-matter multiplets couplings
which generate
 $d=5$ operators. Having in mind valuable way to solve
the mass degeneracy problem we will work on $d=5$ operator generated
proton decay in this scheme. One can see
that what is done in the present paper for the suppression of the nucleon
decay in the framework of $SU(5)$ theory
is analogous to what was proposed for $SO(10)$ SUSY GUT in \cite{gia}
and \cite{zura}.
It is natural to assume, that the  renormalizable couplings
with Higgs multiplets has only the third generation, while the lighter
generations get masses through higher order terms, by the
mixing with there nearest heavy neighbours \cite {fri}; we suppose
that this higher order terms for up quarks are antisymmetric.
Since $qqT$ coupling is symmetric in
the generation indices it vanishes for the light
generations and exists only for the third family. This leads to the strong
suppression of the nucleon decay \cite {zb}.

The paper is organized as follows:
In part 2 we present the $d=5$ operators for general $SU(5)$ theory.
Part 3 deals with proton decay in minimal SUSY SU(5).
Part 4 contains the solution
of $m_l = m_d$ problem in our extended $SU(5)$ SUSY GUTs. In part 5 we
consider proton decay in our
model and part 6 contains discussions and conclusions.

\section{$d=5$ Operators}

Fermion sector of the  SU(5) SUSY GUT consists of the one pair of
fermion supermultiplets $\bar 5+10$ per
generation:
\begin{equation}
\bar 5_{\alpha}=(d^c,~~l)_{\alpha},~~~~~~~~10=(u^c,~~q,~~e^c)_{\alpha}~,
\end{equation}
where $\alpha =1, 2, 3$ is a family index.

The Higgs sector contains the following chiral supermultiplets:
$\Sigma \sim 24$ in adjoint representation of SU(5) and $5$ and $\bar 5$-plets
$H$,  $\bar H$:
\begin{equation}
H=(T,~~H_u),~~~~~~~~\bar H=(\bar T,~~H_d)~.
\end{equation}

The SU(5) invariant Yukawa couplings which
generate masses of the  up and  down quarks and charged leptons
are respectively:
\begin{equation}
10\cdot \hat{\Gamma }_u \cdot 10H,~~~~~~~~~
10\cdot \hat{\Gamma }_d \cdot \bar 5\bar H  , \label {fer}
\end{equation}
where $\hat{\Gamma }_u$ and $\hat{\Gamma }_d$ are Yukawa coupling constants
(family and SU(5) indices are suppressed).
Decomposition of these couplings in general have the form:
\begin{equation}
\Gamma ^u 10\cdot 10\cdot H\to q\hat{Y}_uu^cH_u+q\hat{A}qT+
u^c\hat{B}e^cT~,
\label {dec1}
\end{equation}
\begin{equation}
\Gamma ^d 10\cdot \bar 5\cdot \bar H\to
q\hat{Y}_dd^cH_d+q\hat{C}l\bar T+ u^c\hat{D}d^c\bar
T+e^c \hat{Y}_e lH_d~.
\label {dec2}
\end{equation}

After integrating out the colour Higgses $T$, $\bar T$ with masses
of the order of
$M_{GUT}$ we obtain the following
$d=5$ operators :
\begin{equation}
O_L=\frac {1}{M_{GUT}}(q\hat{A}q)(q\hat{C}l)~,
\label {lrop}
\end{equation}

\begin{equation}
O_R=\frac {1}{M_{GUT}}(u^c\hat{B}e^c)(u^c\hat{D}d^c)~.
\label {00op}
\end{equation}

In general
$\Gamma _{u,d}$ (see (\ref {fer})) can be some
functions of
$\Sigma/M$, where $\Sigma $
breaks SU(5) down to the $G_{123}\equiv U(1)_Y\times SU(2)_W\times
SU(3)_C$ group at the scale $M_{GUT}\simeq 10^{16}$~GeV. $M$ is some fundamental
scale, $M\gg M_{GUT}$.

After diagonalization of the Yukawa matrices by biunitary
transformations:
\begin{equation}
L_u^{+}\hat{Y}_uR_u=\hat{Y}_u^{Diag},~~~~~
L_d^{+}\hat{Y}_dR_d=\hat{Y}_d^{Diag},~~~~~
R_e^{+}\hat{Y}_eL_e=\hat{Y}_e^{Diag}
\label {diag}
\end{equation}
and by proper redefinition of the quark and lepton fields all
operators can be rewritten in the mass eigenstate basis,
where the interaction of the quark-lepton
superfields with colour-triplets have the form:
\begin{equation}
q\hat{A}qT\to q_uL_u^{+}\hat{A}L_d^{*}q_dT~,
\label{qq}
\end{equation}
\begin{equation}
u^c\hat{B}e^cT\to u^cR_u^{T}\hat{B}R_ee^cT~,
\label {ue}
\end{equation}
\begin{equation}
q\hat{C}l\bar T \to q_uL_u^{+}\hat{C}L_el_e\bar T-
q_dL_d^{+}\hat{C}L_el_{\nu }\bar T~,
\label {ql}
\end{equation}
\begin{equation}
u^c\hat{D}d^c\bar T\to u^cR_u^{T}\hat{D}R_dd^c\bar T~;
\label {ud}
\end{equation}
in this basis the current-gauge superfield interactions
have the form:
\begin{equation}
g_2q_u^{+}\hat{V}q_dW^{(+)}+
g_2l_{\nu }^{+}l_eW^{(+)}
\label {cur}
\end{equation}
where $\hat{V}$ is the CKM matrix:
\begin{equation}
\hat{V}=L_u^{T}L_d^{*}
\label {ckm}
\end{equation}

The $O_L$ type $d=5$ operators
which induce proton decay with neutrino emission by exchange
of wino lead to the following baryon number violating four-fermion
interactions
(we omit charge conjugation matrix in fermion braces throughout
this paper):
\begin{equation}
\frac {1}{M_{\tilde{H}_3}}C^{(ud)(d\nu )}_{\delta \alpha \gamma \rho }
\cdot (u^{\delta }_a d^{\alpha }_b)(d^{\gamma }_c\nu ^{\rho })
\varepsilon^{abc}
 \label {ops1}
\end{equation}
where
\begin{equation}
C=C(I)+C(II)+C(III)+C(IV)\;\; ,        \label {sum1}
\end{equation}

\begin{equation}
C^{(ud)(d\nu )}_{\delta \alpha \gamma \rho }(I)=
-g_2^2(L_d^{+}\hat{C}L_e)_{\gamma \rho }
(L_u^{+}\hat{A}L_d^{*})_{\beta \sigma }
V_{\beta \alpha }(V^{+})_{\sigma \delta }
I(\tilde{u^{\beta }},\tilde{d^{\sigma }}) \;\; ,
\label {op1}
\end{equation}

\begin{equation}
C^{(ud)(d\nu )}_{\delta \alpha \gamma \rho }(II)=
-g_2^2(L_d^{+}\hat{A}L_u^{*})_{\gamma \beta }V_{\beta \alpha}
(L_e^{T}\hat{C}^TL_d^{*})_{\rho \sigma }(V^{+})_{\sigma \delta }
I(\tilde{u^{\beta }},\tilde{d^{\sigma }}) \;\; ,
\label {op3}
\end{equation}

\begin{equation}
C^{(ud)(d\nu)}_{\delta \alpha \gamma \rho }(III)=
g_2^2(L_u^{+}\hat{A}L_d^{*})_{\delta \alpha }
(L_u^+\hat{C}L_e)_{\beta \rho}
V_{\beta \gamma}I(\tilde{u}^{\beta },\tilde{e}^{\rho})\;\; ,
\label{350}
\end{equation}

\begin{equation}
C^{(ud)(d\nu )}_{\delta \alpha \gamma \rho }(IV)=
-g_2^2(L_d^{+}\hat{A}L_u^{*})_{\alpha \beta }V_{\beta \gamma }
(L_u^{+}\hat{C}L_e)_{\delta \rho }
I(\tilde{u^{\beta }},\tilde{e^{\rho }})   \;\; ,
\label {op2}
\end{equation}
while the operators with charged leptons are:
\begin{equation}
\frac {1}{M_{\tilde{H_3}}}C^{(ud)(ue )}_{\alpha \beta \delta \rho }
\cdot (u^{\alpha }_a d^{\beta }_b)
(u^{\delta }_ce ^{\rho })
\varepsilon^{abc}      \label {ops2}
\end{equation}
where
\begin{equation}
C=C(I)+C(II)+C(III)+C(IV)\;\; ,        \label {sum2}
\end{equation}

\begin{equation}
C^{(ud)(ue)}_{\alpha \beta \delta \rho }(I)=
-g_2^2(L_u^{+}\hat{A}L_d^{*})_{\alpha \beta}
(L_d^{+}\hat{C}L_e)_{\sigma \rho}\;\;\;
(V^{+})_{\sigma \delta}I(\tilde{d^{\sigma }},\tilde{\nu^{\rho }})\;\; ,
\label{400}
\end{equation}

\begin{equation}
C^{(ud)(ue)}_{\alpha \beta \delta \rho }(II)=
g_2^2(L_u^{+}\hat{A}L_d^{*})_{\alpha \gamma }(V^{+})_{\gamma \delta }
(L_d^{+}\hat{C}L_e)_{\beta \rho }
I(\tilde{d^{\gamma }},\tilde{\nu ^{\rho }}) \;\; ,
\label {op5}
\end{equation}

\begin{equation}
C^{(ud)(ue)}_{\alpha \beta \delta \rho }(III)=
g_2^2(L_u^{+}\hat{C}L_e)_{\delta \rho }
(L_u^{+}\hat{A}L_d^{*})_{\gamma \sigma }V_{\gamma \beta }
(V^{+})_{\sigma \alpha }
I(\tilde{d^{\sigma }},\tilde{u ^{\gamma }})\;\; ,
\label {op4}
\end{equation}

\begin{equation}
C^{(ud)(ue )}_{\alpha \beta \delta \rho }(IV)=
g_2^2(L_u^{+}\hat{A}L_d^{*})_{\delta \gamma }(V^{+})_{\gamma \alpha }
(L_e^{T}\hat{C}^TL_u^{*})_{\rho \omega }V_{\omega \beta }
I(\tilde{d}^{\gamma },\tilde{u^\omega })
\label {op6}
\end{equation}
($\alpha $, $\beta $,...  and  $a, b,...$ are the family and colour indices
respectively).
$C(I)$ and $C(III)$ correspond to the vertex diagrams
(see Figs. 2a,c,e,g)
while $C(II)$ and $C(IV)$ to the box diagrams(see Figs. 2b,d,f,h).  In
(\ref {350})-(\ref {op6})
$I$ denotes the result of the integral over
the loop and is given by the following formula \cite{his}:
\begin{equation}
I(\tilde{u}, \tilde{d})=\frac {1}{16\pi^2}\frac
{m_{\tilde{W}}}{m_{\tilde{u}}^2- m_{\tilde{d}}^2} \left ( \frac
{m_{\tilde{u}}^2}{m_{\tilde{u}}^2- m_{\tilde{W}}^2}\ln \frac
{m_{\tilde{u}}^2}{m_{\tilde{W}}^2}- \frac
{m_{\tilde{d}}^2}{m_{\tilde{d}}^2- m_{\tilde{W}}^2}\ln \frac
{m_{\tilde{d}}^2}{m_{\tilde{W}}^2} \right )~,   \label {int}
\end{equation}
and analogously for $I(\tilde{d}, \tilde{\nu} )$ and
$I(\tilde{u},\tilde{e} )$.

For degenerate squark masses from (\ref {int}) we get:
\begin{equation}
I(\tilde{u},\tilde{d})\to I(m_{\tilde{q}}, x_w)=\frac {1}{16\pi^2}
\frac{\sqrt{x_w}}{m_{\tilde{q}}}
\frac {1}{(1-x_w)^2}
\left(x_w\ln x_w-x_w+1\right)~,
\label {int1}
\end{equation}
where
\begin{equation}
x_w=\left(\frac {m_{\tilde{W}}}{m_{\tilde{q}}}\right)^2~.
\label {x}
\end{equation}

The function $I(m_{\tilde{q}}, x_w)$ has the following  behaviour:
\begin{equation}
16\pi^2 m_{\tilde{q}}I(m_{\tilde{q}}, x_w)=\left\{ \begin{array}{lll}
\frac{\ln x_w}{\sqrt {x_w}} & \mbox{if $x_w \gg 1$} \\
0.5 & \mbox{if $x_w=1$} \\
\sqrt {x_w} & \mbox{if $x_w \ll 1$}
\end{array}
\right.
\label {as}
\end{equation}

In order to estimate stop contribution the following relations
will be used:
\begin{equation}
m^2_{\tilde{t}}=m^2_{\tilde{q}}+m_t^2~,
\label {stopcont}
\end{equation}

\begin{equation}
I(\tilde{t}, \tilde{d})=\frac {1}{16\pi^2}
\frac {m_{\tilde{W}}}{m^2_t}
\frac {m^2_tx_w\ln x_w+(m^2_{\tilde{q}}+m^2_t )
(1-x_w)\ln (1+m^2_t/m^2_{\tilde{q}})}
{m^2_t(1-x_w)+(1-x_w)^2m^2_{\tilde{q}}}~.
\label {int2}
\end{equation}

\section{Proton decay in minimal SUSY SU(5)}

In minimal SU(5) $\Gamma _u$ and $\Gamma _d$ are SU(5)
singlets and
the following relation between the Yukawa matrices occurs
at the GUT scale:
\begin{equation}
\hat{Y}_u=\hat{A}=\hat{B}~,~~~~~~~~
\hat{Y}_d=\hat{Y}_e=\hat{C}=\hat{D}~.
\label {deg}
\end{equation}

In addition in the minimal SU(5) theory $\hat{\Gamma }_u^T=\hat{\Gamma}_u$
which leads to equality of matrices which
are used to transform $u$ and $u^c$ to mass eigenstate basis.
Therefore in the mass eigenstate basis these couplings have the form:
\begin{equation}
\Gamma ^u 10\cdot 10\cdot H\to q\hat{Y}_u^{Diag}u^cH_u+
q_u\hat{Y}_u^{Diag}Vq_dT+
u^c\hat{Y}_u^{Diag}V^*e^cT~,
\label {dec1min}
\end{equation}
$$
\Gamma ^d 10\cdot \bar 5\cdot \bar H\to
q\hat{Y}_d^{Diag}d^cH_d+e^c\hat{Y}_e^{Diag} lH_d+
q_uV^*\hat{Y}_d^{Diag}l_e\bar T-
$$
\begin{equation}
q_d\hat{Y}_d^{Diag}l_{\nu }\bar T
+ u^cV^*\hat{Y}_d^{Diag}d^c\bar T~,
\label {dec2min}
\end{equation}
where V is the CKM matrix.

Taking into account the renormalization effects between
the GUT scale and the SUSY scale the  equalities (\ref {deg}) are violated.
Using the (\ref {deg}) as a boundary conditions, at the
SUSY breaking scale the $\hat{A}\otimes \hat{C}$ and
the $\hat{B}\otimes \hat{D}$
from (\ref {lrop}) can be expressed by the $Y_u\otimes Y_d$ product:
\begin{equation}
\hat{A}\otimes \hat{C}=\left(A_S\right)_L(\hat{Y}_u\otimes \hat{Y}_d)~,
\label {expr1}
\end{equation}

\begin{equation}
\hat{B}\otimes \hat{D}=\left(A_S\right)_R(\hat{Y}_u\otimes \hat{Y}_d)~.
\label {expr2}
\end{equation}
The $A_S$ coefficients describe the renormalization
effect between the GUT and SUSY breaking scales.
The numerical effect of $A_S$ factor will be discussed in what follows.

Calculation of
the amplitude of proton decay consists of two steps:
calculation of 1-loop Feynman diagram(s) where transition between
sparticles and particles occur and calculation of matrix element of
the corresponding four-fermion operator between proton and $K\nu$
system.
Quarks and leptons of the second and/or
the third generations give the main contribution. As these particles
(except neutrinos and $s$-quark) do not participate in proton decay,
their scalar superpartners go into the loop and are transformed into
light species through wino exchange.
Diagrams which describe
the decay $p\to K\nu_{\mu}$ are shown
on Fig. 1. Sum of the amplitudes which are given by diagrams
shown on Fig.
1b and Fig. 1c  equals zero, and we are left with four diagrams shown
on Fig. 1a.

As it usually occurs in SUSY models, vertices on the diagrams
shown on Fig. 1
are known -- they are the same as in nonsupersymmetric SU(5).
Less is
known about propagators -- masses of squarks and wino. Let us remind,
that two Weyl higgsinos from two Higgs doublets with unit charge
mixes with two Weyl winos and two massive Dirac particles are formed.
Mixing matrix contain four parameters $\mu, M_{\tilde{W}}, g_2 v_1/2,
g_2 v_2/2$ which are constrained by one equation:
\begin{equation}
(\frac{g_2 v_1}{2})^2 +(\frac{g_2 v_2}{2})^2 = M_W^2 \;\; .
\label {massw}
\end{equation}

In SUSY GUT $m_{\tilde{W}}=\alpha_2/\alpha_3m_{\tilde{g}}$, where
$m_{\tilde{g}}$ is gluino mass. Since from Tevatron bounds gluino
should weight several hundreds GeV at least, we have
$m_{\tilde{W}} \gg g_2v_1/2, g_2v_2/2$
and instead of dealing with two mass eigenstates in box diagram
we could take into account only $\tilde{W}$-bosino exchange.

Calculating diagrams of Fig. 1a we obtain
\begin{eqnarray}
M &
= & 2\cdot \frac{m_c m_s V_{cd}
V_{us}(g_2)^2}{(v_1/\sqrt{2})\cdot(v_2/\sqrt{2})}
\frac{1}{M_{\tilde{H}_3}}
I(\tilde{u}, \tilde{d})A_S^lA_L
 \times \nonumber \\
&  & [(\nu_L d_L ^a)(u_L ^b s_L ^c) +
(\nu_L s_L ^a)(u_L ^b
d_L ^c)]\varepsilon_{abc} ~,
\label{1}
\end{eqnarray}
where for  triplet higgsino-matter coupling constant
we use $f= m_q V_{ik}/(v/\sqrt{2})$. $W$-bosino
transform sparticles into particles with the
constant $g_2$.
Factors $A$ take into account short ($A_S$) and long ($A_L$) distance
renormalizations of decay amplitude.
The  index $l$ of $A_S^l$ refer to the contribution of the
two light generation particles propagating inside the box diagram
of Fig. 1a . Factor $A_L$ is the long-range renormalization factor
due to the QCD interaction between the SUSY breaking scale and 1~GeV scale
\cite{100}:
\begin{equation}
A_L=\left(\frac{\alpha_3(1 {\rm GeV})}{\alpha_3(m_c)}\right)^{-2/3}
\left(\frac{\alpha_3(m_c)}{\alpha_3(m_b)}\right)^{-18/25}
\left(\frac{\alpha_3(m_b)}{\alpha_3(m_Z)}\right)^{-18/23}
\label{an1}
\end{equation}
and for $\alpha_3(m_Z)=0.120$ using $\alpha_3$ running at
two loops we get $A_L=0.32$.

In (\ref{1}) factor 2 comes from 2 diagrams, $V_{ik}$ are the
elements of Kobayashi-Maskawa matrix, $v_1$ and $v_2$ are Higgs fields
vacuum
expectation values and $M_{\tilde{H}_3}$ is mass of Higgs triplet,
$M_{\tilde{H}_3} \approx M_{GUT} = 10^{16}$~GeV (let us remind that
for minimal SUSY SU(5) Higgs triplets interactions with quarks and
leptons which generate operator $O_L$ are described by the
Kobayashi-Maskawa matrix).

It is convenient to rewrite (\ref {1}) introducing angle $\beta $,
$\tan \beta =v_1/v_2$ and expressing $v_1^2+v_2^2$ through $G_F$:
\begin{equation}
M = \frac{8\sqrt{2} G_F m_c m_s g_2^2 V_{cd} V_{us}}
{M_{\tilde{H}_3} \sin(2\beta)}
I(m_{\tilde{q}}, x_w)A_LA_S^l
[(\nu_L d_L^a)(u_L^b s_L^c)
+(\nu_L s_L^a)
(u_L^b d_L^c)]\varepsilon_{abc}~.
\label{2}
\end{equation}

For the matrix element of operator (\ref{2}) between
hadronic states we use the result obtained in \cite{5} :
\begin{equation}
\langle \nu
 K^+|\varepsilon^{abc}[(\nu_L d_L^a)(u_L^b s_L^c)+(\nu_L
s_L^a)(u_L^b d_L^c)]|p\rangle  =
\frac{\sqrt{2}\tilde{\beta } G}{(M_{\Lambda}+M_{\Sigma})/2}
(\nu P_Lp)K ~,
\label{110}
\end{equation}
where $P_L=\frac{1}{2}(1+\gamma_5)$.

Finally, from (\ref{2}) and (\ref{110}) we get:
\begin{equation}
M_{p\to K\nu} =\frac{16\cdot \tilde{\beta }
G G_F m_c m_s g_2^2 V_{cd} V_{us}}
{M_{\tilde{H}_3}\sin 2
\beta [(M_{\Lambda}+M_{\Sigma})/2]}
I(m_{\tilde{q}}, x_w)
A_LA_S^l (\nu P_Lp)\equiv
x(\nu P_Lp)K ~.
\label{15}
\end{equation}

Short distance renormalization  factor $A_S$ depends on the numerical
value of $\tan \beta $ \cite{his}. Making an attempt to suppress
the proton decay amplitude  we take the value
of $\tan \beta $ which minimize the ratio $\frac{A_S^l}{\sin 2\beta}$;
so we use $\sin \beta= 0.965$ ($\tan \beta =3.68$), $A_S^l=1.4$,
$\frac{A_S^l}{\sin 2\beta}=2.79$ (detailed calculations will be
published in an extended paper).

Substituting numbers in (\ref{15})  we get:
$$
\Gamma =\frac{(m_p^2
-m_K^2)^2}{32\pi m_p^3}x^2 =\left(\frac{m_c}{1.3~ {\rm GeV}}
\frac{m_s}{175~ {\rm MeV}}
\frac{10^{16} {\rm GeV}}{M_{\tilde{H}_3}}\frac{\tilde{\beta }}
{0.007 {\rm GeV}^3}
\right. \times
$$
\begin{equation}
\frac{I(m_{\tilde{q}}, x_w)}{I(500 {\rm GeV}, 1)}
\frac{A_L}{0.32}
\frac{A_S^l}{1.4}\frac{0.51}{\sin 2\beta }
\left.
\right)^2
\frac{1}{4.5\cdot 10^{27} {\rm years}}~.
\label{16}
\end{equation}

Modern
experimental bound is $\tau_{p\to K\nu} > 10^{32}$ years \cite{6}.
Variation of parameters could hardly help in enhancing proton lifetime
that much.
$\tilde{q}$ and $\tilde{W}$ with mass scale several
TeV did not seem appealing, neither is $m_{\tilde{H}_3} \approx
10^{18}$~GeV (let us remind that SUSY GUT unification scale is
$M_{GUT} = 10^{16}$~GeV and $M_{\tilde{H}_3} \approx \lambda/g \cdot
M_{GUT}$, where $g$ is gauge coupling at unification scale, while
$\lambda $ is a constant of Higgs multiplets selfinteraction).

\begin{table}
\caption{The value of proton life time in Standard SUSY $SU(5)$
in units of
$4.5\cdot 10^{27}$~yeares.
Allowed domain of $m_{\tilde{W}}$,
$m_{\tilde{q}}$ values is in low left corner. }

\label{t:minsu5}
$$\begin{array}{|c|c|c|c|c|c|c|}
\hline
m_{\tilde{W} } & & & & & & \\
&100~ {\rm GeV} &200~ {\rm GeV} &500~ {\rm GeV} &1~ {\rm TeV}
&5~ {\rm TeV} &10~ {\rm TeV}  \\
m_{\tilde{q} }& & & & & & \\
\hline
\hline
& & & & & & \\
100~ {\rm GeV}&0.04 &0.03 &0.04 &0.07 &0.54 &1.48  \\
& & & & & & \\
\hline
& & & & & & \\
200~ {\rm GeV}&0.31 &0.16 &0.13 &0.17 &0.84 &2.14  \\
& & & & & & \\
\hline
& & & & & & \\
500~ {\rm GeV}&7.7 &2.6 &1 &0.78 &1.8 & 4.0 \\
& & & & & & \\
\hline
& & & & & & \\
1 ~{\rm TeV}&1.1\cdot 10^2 &30.7 &7.8 &4 &4.2 &7.4  \\
& & & & & & \\
\hline
& & & & & & \\
5~ {\rm TeV}&6.3\cdot 10^4 &1.6\cdot 10^4 &2.7\cdot 10^3
&7.7\cdot 10^2 & 10^2 &78  \\
& & & & & & \\
\hline
& & & & & & \\
10~ {\rm TeV}& 10^6 &2.5\cdot 10^5 &4.1\cdot 10^4
&1.1\cdot 10^4 &7.8\cdot 10^2 & 4\cdot 10^2 \\
& & & & & & \\
\hline
\end{array}$$
\end{table}

The proton lifetime  for the different
values of squark and wino masses
are presented in Table \ref{t:minsu5} .
As we see the reasonable lifetime is obtained for $m_{\tilde{q}}=5-10$~TeV
and $m_{\tilde{W}}$  about 1~TeV or less ( in this domain
$x_w$ is  small and the function
$m_{\tilde{q}}I(m_{\tilde{q}}, x_w)$ can be described by the asymptotic
formula (\ref {as})).
If one wants to have lighter quarkino, with mass less then, say 1~TeV,
then proton decay should be somehow suppressed.

Before we will go to the main part of our paper let us estimate how
much the contribution of the third generation particles in the proton
decay amplitude can be. If instead of $\tilde{s}_L$ ($\tilde{\mu}_L$)
on Fig. 1a we substitute $\tilde{b}_L$ ($\tilde{\tau}_L$), we will get
the following extra factor in the amplitude (\ref{1}):
\begin{equation}
\frac{A_S^h  (\tilde{b})}{A_S^l }
\frac{m_b}{m_s} |\frac{V_{ub}}{V_{us}}|=
\frac{4.1\div 4.5}{0.1\div
0.3}\times \frac{0.002 \div 0.005}{0.22} =0.1 \div 1 \;\; .
\label{17}
\end{equation}

$A_S^h$ is the short range renormalization factor for the heavy
generations and
$A_S^h(\tilde{b})=A_S^l$.

Stop substituted instead of $\tilde{c}_L$ on the upper line of Fig.
1a lead to the following factor in the amplitude (\ref{1}):
$$
\frac{A_S^h(\tilde{t})}{A_S^l}
\frac{I(\tilde{t}, \tilde{d})}{I(\tilde{u}, \tilde{d})}
\frac{\eta_t m_t}{m_c}|\frac{V_{td}V_{ts}}{V_{cd}}| =
$$
\begin{equation}
\frac{1.9}{1.4}\frac{6.1\cdot 10^{-6}}{6.3\cdot 10^{-6}}
\frac{2.4\cdot 180}{1\div
1.6}\frac{(0.004 \div 0.014)\times (0.034 \div 0.046)}{0.22} =
0.3 \div 1.5~,
\label{18}
\end{equation}
where we use $m_{\tilde{q}}=m_{\tilde{W}}=500$~GeV.
$A_S^h(\tilde{t})=1.9$.

Let us stress that amplitude (\ref{2}) is defined at $\mu =1$~GeV.
Since t-quark mass is not renormalized from the virtuality
which equals to its pole
value $m_t=180$~GeV to virtuality 1~GeV a compensation factor $\eta_t$
should be introduced in $A_L:$
\begin{equation}
\eta_t =\frac{\bar m_t(1 \; \mbox{\rm GeV})}{m_t}  =
\left[\frac{\alpha_3(1 \; \mbox{\rm
GeV})}{\alpha_3(m_t)}\right]^{\frac{4}{11-\frac{2}{3}\cdot 5}}
\label{19}
\end{equation}
for $\alpha_3(M_Z)=0.120$  this factor equals $2.4$.

From (\ref{17}) and
(\ref{18}) we see, that for the maximum mixing between first and third
generation allowed
experimentally contribution of third generation particles
into proton decay can compete with that of second generation. Amplitude with
intermediate stop interfere with that with intermediate scalar charm quark
and may partly cancel it; however compensation with 1\% accuracy which is
needed to satisfy experimental bound $\tau_{p\to\nu K} > 10^{32}$ years is
unnatural.

\section{Predictive ansatz for Yukawa couplings and suppression of
proton decay}

By focusing on the fermion mass pattern, it is natural to suggest that
only the third, heaviest family acquires masses through ordinary
renormalizable Yukawa couplings, while the mass terms of other families
appear from higher order (may be Planck scale) operators, which can be
generated by heavy particle exchange mechanism \cite{18}:

\begin{equation}
W_Y^u=\frac {1}{4}C'\cdot 10\cdot 10H+
\frac {1}{4}\frac {B'}{M}\cdot 10\cdot 10\cdot \Sigma H+
\frac {A'}{M^2}\cdot 10\cdot 10\cdot \Sigma^2 H~,
\label {up}
\end{equation}
\begin{equation}
W_Y^d=\sqrt 2d\cdot 10\cdot \bar 5 \bar H+
\sqrt 2\frac {b'}{M}\cdot 10\cdot \bar 5 \cdot \Sigma \cdot \bar H+
\sqrt 2\frac {a'}{M^2}\cdot 10\cdot \bar 5 \cdot \Sigma^2 \cdot \bar H~,
\label {down}
\end{equation}
where $C'$, $B'$,... are matrices in generation space.

In order to be
closer to the realistic mass matrices let us suggest
for them the following form:
\begin{eqnarray}
C'_{\alpha \beta }\sim \delta_{3\alpha }\delta_{3\beta }~,~~~~~
B_{\alpha \beta }'\sim \delta_{3\alpha }\delta_{2\beta }+
k_B\delta_{2\alpha }\delta_{3\beta }~,~~~~~
A_{\alpha \beta }'\sim \delta_{2\alpha }\delta_{1\beta }+
k_A\delta_{1\alpha }\delta_{2\beta } \nonumber ~,\\
d_{\alpha \beta }\sim \delta_{3\alpha }\delta_{3\beta }~,~~~~~
b'_{\alpha \beta }\sim \delta_{2\alpha }\delta_{2\beta }~,~~~~~
a_{\alpha \beta }'\sim \delta_{2\alpha }\delta_{1\beta }+
k_a\delta_{1\alpha }\delta_{2\beta }~.
\label {ans}
\end{eqnarray}

In (\ref {ans}) $k$ are Clebsch factors.
For $W_{Y}^u$ we get:
\begin{equation}
10\times 10=\bar 5+\overline {45}+\overline {50} \;\; ,
\label{480}
\end{equation}
while for $W_Y^d$ we have:
\begin{equation}
10\times\bar{5}=5+45 \;\; .
\label {prod1}
\end{equation}
For the bilinear Higgs fields product we have:
\begin{equation}
24 \times 5 = 5+45+70 \;\; ,
\label{481}
\end{equation}
so the $\Sigma H$ could belong to 5 or 45 and in these cases $B'$ is
symmetric ($k_B=1$) or antisymmetric ($k_B=-1$), respectively.
However, because in $24\times 24\times 5$ several invariants
of 5 and 45 plets and also 50-plet do occur, there exist many
invariants and many possibilities for $k_A$ and  $k_a$'s.

In what follows matrices $A'$, $B'$ will be taken antisymmetric
($k_A=k_B=-1$)
and this will be crucial for the proton stability.

In other
words we suppose that for some reason the
composite operators $\Sigma H$ and
$\Sigma^2 H$ are participate
in expression (\ref {up}) only
in representation $45$ .

Insertion of $\Sigma $ in higher order terms helps to
avoid the  degeneracy  of the masses
of down quarks and charged leptons.
As it was assumed
matrices $B'$ and $A'$ are antisymmetric, while $a'$ is symmetric
with respect to the family indices. (This can be attributed to
some symmetry reasons).
Then after substituting the VEVs of $\Sigma $,
$H$ and $\bar H$  Yukawa matrices for up and down
quarks and leptons will have the forms:
\begin{equation}
\begin{array}{ccc}
\hat{Y}_u=~~ \\
\end{array}
\hspace{-6mm}\left(
\begin{array}{ccc}
0& A & 0 \\
-A& 0& B \\
0& -B& C \end{array}
\right),~~~~
\begin{array}{ccc}
\hat{Y}_d=~~ \\
\end{array}
\hspace{-6mm}\left(
\begin{array}{ccc}
0& a_1 & 0 \\
a_1& b_1& 0 \\
0& 0& d \end{array}
\right),~~~~
\begin{array}{ccc}
\hat{Y}_e=~~ \\
\end{array}
\hspace{-6mm}\left(
\begin{array}{ccc}
0& a_2 & 0 \\
a_2& b_2& 0 \\
0& 0& d \end{array}
\right)~.
\label {matUDE}
\end{equation}

Because according to our choice  $B'$ and $A'$ matrices are
antisymmetric in the family space while  the $qqT$
coupling is symmetric on the generation indices only 33
element of the matrix $\hat{A}$ is nonvanishing (as we will see this
fact is crucial for proton decay):
\begin{equation}
\begin{array}{ccc}
\hat{A}=~~ \\
\end{array}
\left(
\begin{array}{ccc}
0&0 & 0 \\
0& 0& 0 \\
0& 0&C \end{array}
\right)~,
\nonumber \\
\begin{array}{ccc}
\hat{B}=~~ \\
\end{array}
\hspace{-6mm}\left(
\begin{array}{ccc}
0& \tilde{A} & 0 \\
-\tilde{A}& 0 & \tilde{B} \\
0&-\tilde{B} & C \end{array}
\right)~,
\label {AB}
\end{equation}

\begin{equation}
\begin{array}{ccc}
\hat{C}=~~ \\
\end{array}
\hspace{-6mm}\left(
\begin{array}{ccc}
0&\tilde{a}_1 & 0 \\
\tilde{a}_1& \tilde{b}_1& 0 \\
0& 0&d \end{array}
\right),~~~~
\nonumber \\
\begin{array}{ccc}
\hat{D}=~~ \\
\end{array}
\hspace{-6mm}\left(
\begin{array}{ccc}
0& \tilde{a}_2 & 0 \\
\tilde{a}_2& \tilde{b}_2& 0 \\
0& 0& d \end{array}
\right)~.
\label {CD}
\end{equation}

The values of matrix elements of matrices
(\ref {matUDE}), (\ref {AB}), (\ref {CD})
depend on the $SU(5)$ representations to which higher order Higgs
terms in (\ref {up}), (\ref {down}) belong. In numerical estimates
we will take $\tilde{a_1}=a_1$; concerning $\tilde{b_1}$ two possibilities
will be considered (see later).

Structure of the matrices (\ref {matUDE}) resembles
the ansatz proposed by Georgi and Jarlskog
in an $SU(5)$ GUT \cite{20}.
Lately a number of authors \cite {22,23}
have reexamined this texture in a supersymmetric context.
From (\ref {matUDE}) it is easy to find, that

$$
A \approx\sqrt {\lambda _u \lambda _c}~,~~~
B \approx\sqrt {\lambda _c \lambda _t}~,~~~
C \approx \lambda _t~,~~~
a_1 \approx \sqrt {\lambda _d \lambda _s}~,~~~
b_1 \approx \lambda _s~,~~~
$$
\begin{equation}
a_2 \approx\sqrt {\lambda _e \lambda _{\mu }}~,~~~
b_2 \approx\lambda _{\mu }~,~~~
d=\lambda _b=\lambda _{\tau }~.    \label {cons}
\end{equation}
The Yukawa matrices are diagonalized by the transformations
given in (\ref {diag}), where for $\hat{Y}$ from
(\ref {matUDE}) we have:
\begin{equation}
L_u=L_u^{(23)}\cdot L_u^{(12)}~,~~~~
L_d=L_d^{(12)}~,~~~~
L_e=L_e^{(12)}~,
\label {blok}
\end{equation}
where
\begin{equation}
\nonumber \\
\begin{array}{ccc}
L_u^{(12)}=~~ \\
\end{array}
\hspace{-6mm}\left(
\begin{array}{ccc}
\cos \theta_{12}^{L_u}&-\sin \theta_{12}^{L_u} & 0 \\
\sin \theta_{12}^{L_u}&\cos \theta_{12}^{L_u} &0\\
0&0 & 1 \end{array}
\right),~~~~
\nonumber \\
\begin{array}{ccc}
L_u^{(23)}=~~ \\
\end{array}
\hspace{-6mm}\left(
\begin{array}{ccc}
1& 0& 0 \\
0& \cos \theta_{23}^{L_u}& -\sin \theta_{23}^{L_u} \\
0& \sin \theta_{23}^{L_u}& \cos \theta_{23}^{L_u} \end{array}
\right),~~~~
\label {bloku}
\end{equation}

\begin{equation}
\nonumber \\
\begin{array}{ccc}
L_d^{(12)}=~~ \\
\end{array}
\hspace{-6mm}\left(
\begin{array}{ccc}
\cos \theta_{12}^{L_d}&-\sin \theta_{12}^{L_d} & 0 \\
\sin \theta_{12}^{L_d}&\cos \theta_{12}^{L_d} &0\\
0&0 & 1 \end{array}
\right),~~~~
\nonumber \\
\begin{array}{ccc}
L_e^{(12)}=~~ \\
\end{array}
\hspace{-6mm}\left(
\begin{array}{ccc}
\cos \theta_{12}^{L_e}& -\sin \theta_{12}^{L_e}& 0 \\
\sin \theta_{12}^{L_e}& \cos \theta_{12}^{L_e}& 0 \\
0& 0& 1 \end{array}
\right),~~~~
\label {blokdl}
\end{equation}

\begin{equation}
\sin \theta_{12}^{L_u}\simeq -\frac {AC}{B^2}~,~~~
\sin \theta_{23}^{L_u}\simeq -\frac {B}{C}~,~~~
\sin \theta_{12}^{L_d}\simeq -\frac {a_1}{b_1}~,~~~
\sin \theta_{12}^{L_e}\simeq -\frac {a_2}{b_2}~.
\label {angles}
\end{equation}
Therefore at the GUT scale we have:
\begin{equation}
\nonumber \\
\begin{array}{ccc}
L_u=~~ \\
\end{array}
\hspace{-6mm}\left(
\begin{array}{ccc}
1& \frac {AC}{B^2} & 0 \\
-\frac {AC}{B^2}& 1& \frac {B}{C} \\
\frac {A}{B}&-\frac {B}{C} & 1 \end{array}
\right),~~~~
\nonumber \\
\begin{array}{ccc}
L_d=~~ \\
\end{array}
\hspace{-6mm}\left(
\begin{array}{ccc}
1& \frac {a_1}{b_1} & 0 \\
-\frac {a_1}{b_1}& 1& 0 \\
0& 0& 1 \end{array}
\right),~~~~
\nonumber \\
\begin{array}{ccc}
L_e=~~ \\
\end{array}
\hspace{-6mm}\left(
\begin{array}{ccc}
1& \frac {a_2}{b_2} & 0 \\
-\frac {a_2}{b_2}& 1& 0 \\
0& 0& 1 \end{array}
\right)~,
\label {Lude}
\end{equation}
while for $R_{u,d,e}$ matrices we have:
\begin{equation}
\nonumber \\
\begin{array}{ccc}
R_u=~~ \\
\end{array}
\hspace{-6mm}\left(
\begin{array}{ccc}
1& \frac {AC}{B^2} & 0 \\
-\frac {AC}{B^2}& 1& -\frac {B}{C} \\
-\frac {A}{B}&\frac {B}{C} & 1 \end{array}
\right),~~~~
\nonumber \\
\begin{array}{ccc}
R_d=~~ \\
\end{array}
%&
\hspace{-6mm}\left(
\begin{array}{ccc}
-1& \frac {a_1}{b_1} & 0 \\
\frac {a_1}{b_1}& 1& 0 \\
0& 0& 1 \end{array}
\right),~~~~
%&~~~~
\nonumber \\
\begin{array}{ccc}
R_e=~~ \\
\end{array}
\hspace{-6mm}\left(
\begin{array}{ccc}
-1& \frac {a_2}{b_2} & 0 \\
\frac {a_2}{b_2}& 1& 0 \\
0& 0& 1 \end{array}
\right)~.
\label {Rude}
\end{equation}

After diagonalization the Yukawa matrices have the form:
\begin{equation}
\hat{Y}_u^{Diag}=(\lambda _u,~\lambda _c,~\lambda _t)~,~~
\hat{Y}_d^{Diag}=(\lambda _d,~\lambda _s,~\lambda _b)~,~~
\hat{Y}_e^{Diag}=(\lambda _e,~\lambda _{\mu },~\lambda _{\tau })~.
\label {dude}
\end{equation}

From (\ref {ckm}), (\ref {cons}) and (\ref {Lude}) one can find
the CKM matrix elements:
$$
V_{us}=\sqrt {\frac {\lambda _d}{\lambda _s}}-
\sqrt {\frac {\lambda _u}{\lambda _c}}~,~~~~
V_{cb}=-\sqrt {\frac {\lambda _c}{\lambda _t}}~,~~~
V_{ub}=\sqrt {\frac {\lambda _u}{\lambda _t}}~,
$$
\begin{equation}
V_{ts} =\sqrt{\frac{\lambda_c}{\lambda_t}}~, ~~~~
V_{td} =-\sqrt{\frac{\lambda_c \lambda_d}{\lambda_t \lambda_s}}~.
\label {V}
\end{equation}

As we see on the GUT scale the value of the $V_{cb}$ element
is too large ($V_{cb}^{exp} =0.036 \div 0.046$).
It appears \cite{22}-\cite{23},
that the desirable relations between masses and mixing angles are
obtained on the electroweak breaking scale after
taking into account  the
renormalization effects.

\section{Proton decay in extended SUSY SU(5)}

Let us estimate now the proton decay probability in our model.
Let us start
with $p\to K^+\nu_{\mu }$ mode, which dominates in the minimal SU(5).
This decay is described by the diagrams  Fig. 2a-d.

From (\ref {Lude}), (\ref {AB}) it is easy to see,
that $(L_d^{+}\hat{A}L_u^{*})_{\alpha \beta }$
exactly vanish for $\alpha =1, 2$. Therefore, the amplitudes
(\ref {op3}), (\ref {350}) and (\ref {op2}) do not lead
to the proton decay as they produce
$b$-quark in final state.

The amplitude described by eq. (\ref {op1})
is suppressed
for another reason: as we see from (\ref {op1}) this amplitude do not
vanish if $\sigma =3$. However
in the inner line of diagram  Fig.2a $\tilde{u}$, $\tilde{c}$ and $\tilde{t}$
squarks run.
Assuming for a moment that integral $I$ in (\ref {op1})
is family independent, taking sum over $\beta $  and using (\ref {ckm})
we see that (\ref {op1}) is proportional to
$(L^+_dAL_d^*)_{\alpha 3}$ and in the external line still  the $b$
quark is produceed.
In this way we see that $d$ or $s$ quarks which can
participate in proton decay are not emitted. However the above
argument is valid only if the equality
$m_{\tilde{u}}=m_{\tilde{c}}=m_{\tilde{t}}$
holds; heaviness of the top quark breaks last equality, so $p\to K\nu$
decay through diagram Fig. 2a do occur.
So, taking into account the shift of $I$ function the nucleon decay
will take place due to heavy stop exchange, but the suppression factor
$\frac {\Delta I}{I}$ will appear, where $\Delta I$ is:
\begin{equation}
\Delta I=I(\tilde{t},\tilde{b})-I(\tilde{u}^{\beta },\tilde{b}),~~~~~~
\beta =1, 2
\label {dint}
\end{equation}

These arguments work for both $p\to K\nu $ and $p\to \pi \nu $
decays.

Introducing the parameter
\begin{equation}
F_w=\frac{8g_2^2GG_FA_S^h(\tilde{t})
A_L\tilde{\beta } I}{M_{\tilde{H_3}}\sin 2\beta }~,
\label{fw}
\end{equation}
for the  $p\to K\nu^{\alpha }$ decay widths
\footnote{necessary matrix elements are presented in the end of this section}
we get:
$$
\Gamma (p\to K\nu_{\mu })=\frac{(m_p^2-m_K^2)^2}{32\pi m_p^3}
F_w^2\left(\frac {\Delta I}{I}\right)^2
\left(\frac {v_2}{\sqrt 2}\right)^2\bar m_t^2|V_{ub}|^2\times
$$
\begin{equation}
\left|V_{ts}(C_{12}+C_{22}L^+_{d12})\frac{2/3\alpha }
{(M_{\Lambda}+M_{\Sigma})/2}+
V_{td}C_{22}\frac{1-2/3\alpha }{M_{\Lambda }}\right|^2~,
\end{equation}
$$
\Gamma (p\to K\nu_e)=\frac{(m_p^2-m_K^2)^2}{32\pi m_p^3}
F_w^2\left(\frac {\Delta I}{I}\right)^2
\left(\frac {v_2}{\sqrt 2}\right)^2\bar m_t^2|V_{ub}|^2\times
$$
$$
\left|V_{ts}(C_{12}(L_e)_{21}+C_{21}(L_d^+)_{12}+
C_{22}(L_d^+)_{12}(L_e)_{21})\frac{2/3\alpha }{(M_{\Lambda}+M_{\Sigma})/2}+
\right.
$$
\begin{equation}
\left.
V_{td}(C_{21}+C_{22}(L_e)_{21})\frac{1-2/3\alpha }{M_{\Lambda }}\right|^2~,
\end{equation}
while the widths of the $p\to \pi \nu^{\alpha}$ decays are:
\begin{equation}
\Gamma (p\to \pi \nu_{\mu })=\frac{m_p}{32\pi m_n^2}
F_w^2\left(\frac {\Delta I}{I}\right)^2
\left(\frac {v_2}{\sqrt 2}\right)^2\bar m_t^2|V_{ub}V_{td}|^2
\left|C_{12}+C_{22}(L_d^+)_{12}\right|^2~,
\end{equation}
$$
\Gamma (p\to \pi \nu_e)=\frac{m_p}{32\pi m_n^2}
F_w^2\left(\frac {\Delta I}{I}\right)^2
\left(\frac {v_2}{\sqrt 2}\right)^2\bar m_t^2|V_{ub}V_{td}|^2\times
$$
\begin{equation}
\left|C_{12}(L_e)_{21}+C_{21}(L_d^+)_{12}+
C_{22}(L_e)_{21}(L_d^+)_{12}\right|^2~,
\end{equation}
where $\bar{m_t}=2.4\cdot 180$~GeV (an artifact of the $A_L$
definition).

\begin{table}\caption{An order of magnitude estimates of the proton partial
life times in the units of $10^{32}$ years.}
\label{t:ourmodel}
$$\begin{array}{|c|c|c|}
\hline & C_{22}=0 & C_{22}=2\lambda_s \\\hline
 p\to K \nu_{\mu } &\sim 10 &\sim 3 \\
p\to K \nu_e & \sim 200& \sim 200  \\
p\to \pi \nu_{\mu } & \sim 10 &\sim 10 \\
p\to \pi \nu_e & \sim 300& \sim 400  \\
\hline p\to K^0\mu ^+ & \sim 10 & \sim 0.01 \\
p\to K^0e ^+ & \sim 1 &\sim 1\\
p\to {\pi}^0\mu ^+ & \sim 0.05  & \sim 0.05  \\
p\to {\pi}^0e ^+ & \sim 1& \sim 1 \\
\hline
\end{array}$$
\end{table}

Crucial for the suppression of the $p\to K\nu $ decay mode is the form of
matrix $A$. We had study the renormalization of the
matrix $A$ from GUT to the SUSY breaking scale and it
appears that its form  is not changed,
so the results presented in this
section are valid also for the case when the renormalization effects are
taken into account.

Proton decays $p\to Kl^+$
and $p\to \pi e^+$ in our model are described by the box
diagrams shown on Fig. 2f and 2h and the
vertex diagram shown on Fig. 2g amplitude of which is proportional to
$\Delta I/I$
for the same reason as that described by the
diagram of Fig. 2a (see the beginning
of this section).
Vertex diagram shown on Fig. 2e produces $b$
quark, so it is irrelevant for proton decay.
For proton decay widths we obtain:
$$
\Gamma (p\to K \mu )=\frac{(m_p^2-m_K^2)^2}{32\pi m_p^3}
\left(\frac{1-2\alpha}{ M_{\Sigma }}\right)^2F_w^2
\left(\frac {v_2}{\sqrt 2}\right)^2\bar m_t^2
|V_{ub}|^2\left|\sqrt {\frac{m_u}{m_t}}
\left(2C_{22}+ V_{us}C_{12} +
\right.
\right.
$$
\begin{equation}
2C_{21}(L_e)_{12}+C_{12}(L_d^+)_{21}
\left.+C_{12}(L_u^+)_{21}\right)
-\left(\frac {\Delta I}{I}\right)
\left.V_{ts}\left(C_{12}+C_{22}(L_u^+)_{12}\right)\right|^2~,
\end{equation}
$$
\Gamma (p\to K e )=\frac{(m_p^2-m_K^2)^2}{32\pi m_p^3}
\left(\frac{1-2\alpha}{ M_{\Sigma }}\right)^2F_w^2
\left(\frac {v_2}{\sqrt 2}\right)^2\bar m_t^2
|V_{ub}|^2
\left|2\sqrt {\frac{m_u}{m_t}}\right.
\left(C_{21}+
\right.
$$
\begin{equation}
\left.
C_{22}(L_e)_{21}\right)-
\left(\frac {\Delta I}{I}\right)
V_{ts}
\left(C_{12}(L_e)_{21}+C_{21}(L_u^+)_{12}+
\left.C_{22}(L_u^+)_{12}(L_e)_{21}\right)\right|^2~,
\end{equation}
$$
\Gamma (p\to \pi \mu )=\frac{1}{64\pi m_p}F_w^2
\left(\frac {v_2}{\sqrt 2}\right)^2\bar m_t^2|V_{ub}|^2\times
$$
\begin{equation}
\left|\sqrt {\frac{m_u}{m_t}}
\left(2C_{12}+C_{22}V_{cd}\right)-
\left(\frac {\Delta I}{I}\right)
V_{td}\left(C_{12}+C_{22}(L_u^+)_{12}\right)\right|^2~,
\end{equation}
$$
\Gamma (p\to \pi e )=\frac{1}{64\pi m_p}F_w^2
\left(\frac {v_2}{\sqrt 2}\right)^2\bar m_t^2
|V_{ub}|^2\times
$$
$$
\left|2\sqrt {\frac{m_u}{m_t}}
\left(C_{12}(L_e)_{21}+C_{21}(L_d^+)_{12}+C_{22}(L_d^+)_{12}(L_e)_{21}\right)-
\right.
$$
\begin{equation}
\left.
\left(\frac {\Delta I}{I}\right)
V_{td}
\left(C_{12}(L_e)_{21}+C_{21}(L_u^+)_{12}+C_{22}(L_u^+)_{12}(L_e)_{21}\right)
\right|^2~.
\end{equation}

The desirable at GUT scale relations
$\frac {\lambda_{\mu} }{\lambda_s}=3$ or $-3$ in our model occur
for $C_{22}\equiv \tilde{b}_1=0$ or $C_{22}=2\lambda_s$.
In the first case the strong
suppression of the $p\to K\mu$ mode will occur. In numerical
estimates we  consider both these  cases.

Proton partial lifetimes for   $C_{22}=2\lambda_s$
and $C_{22}=0$ and  for the values of the parameters
from Table \ref{t:tabpar}
are presented in Table \ref{t:ourmodel} .
As we see proton partial lifetimes with emission of neutrino
for values $(m_{\tilde{q}}, m_{\tilde{W}})=(500~{\rm GeV}, 500~{\rm GeV})$
in both cases $C_{22}=2\lambda_s$ and $C_{22}=0$ are
compatible with the experimental
data. For the case $C_{22}=2\lambda_s $ the
decays $p\to K\mu$ and $p\to \pi \mu$ are too fast and we have to change
the masses of SUSY particles. For example for
$(m_{\tilde{q}}, m_{\tilde{W}})=(1~{\rm TeV}, 100~{\rm GeV})$
we get
$\tau (p\to K\mu)=10^{32}$~years and
$\tau (p\to \pi \mu)=10^{33}$~years.For the case $C_{22}=0$
 the $p\to \pi \mu$
mode dominates and for
$(m_{\tilde{q}}, m_{\tilde{W}})=(600~{\rm GeV}, 100~{\rm GeV})$~
$\tau (p\to \pi \mu)=10^{32}$~years.

At the end of this section let us present the results of
the calculation of the matrix elements which contribute into
proton decay in our model:
\begin{equation}
\langle \nu
 K|(us)(\nu d)|p\rangle
=\frac{\tilde{\beta } G}{(M_{\Lambda}+M_{\Sigma})/2}
\frac{2\sqrt{2}\alpha }{3}
(\nu P_Lp)K \;\; ,
\label{c1}
\end{equation}
\begin{equation}
\langle \nu
 K|(ud)(\nu s)|p\rangle =\frac{\tilde{\beta } G}{M_{\Lambda }}
\sqrt{2}(1-2/3\alpha )(\nu P_Lp)K \;\; ,
\label{c2}
\end{equation}
\begin{equation}
\langle \nu
 \pi |(ud)(d\nu )|p\rangle =\frac{\tilde{\beta } G}{M_n}
\sqrt{2}(\nu P_Lp)\pi \;\; ,
\label{c3}
\end{equation}
\begin{equation}
\langle l
 K|(us)(u l)|p\rangle =\frac{\tilde{\beta } G}{M_{\Sigma}}
\sqrt{2}(1-2\alpha) (l P_Lp)K \;\; ,
\label{c4}
\end{equation}
\begin{equation}
\langle l
 \pi|(ud)(u l)|p\rangle =\frac{\tilde{\beta } G}{M_n}
(l P_Lp)\pi ~.
\label{c5}
\end{equation}

\begin{table}
\caption{Numerical values of parameters which were used in
the estimates of
the proton partial lifetimes.}
\label{t:tabpar}
$$\begin{array}{|c|c|c|c|c|c|c|}
\hline
m_{\tilde{W}}&m_{\tilde{q}}&m_u&m_d&m_s&
m_c&m_t\\
\hline
500~{\rm GeV}&500~{\rm GeV}&3.6~{\rm MeV}&6~{\rm MeV}&151~{\rm MeV}&
1.3~{\rm GeV}&180~{\rm GeV}\\
\hline
\hline
V_{ub}&V_{ts}&V_{td}&\sin 2\beta &A_S^h(\tilde{t})&
M_{\tilde{H_3}} &\alpha_3 (M_Z)
 \\
\hline
0.003&0.05&0.01&0.51&1.9&
10^{16}~{\rm GeV}&0.120 \\
\hline
\end{array}$$
\end{table}

\section{Discussions}

One of the most appealing next step after minimal standard
$SU(3)\times SU(2)\times U(1)$ model is Grand Unification Theory.
Both theoretical argument (hierarchy problem) and experimental
measurements (values of electroweak mixing angle and $\alpha_s$)
prefer, select or point out on SUSY GUT. Simplest variant is SU(5)
SUSY GUT. However, minimal version of the model has two drawbacks:
too short proton life time and famous ratio: $m_d/m_s=m_e/m_{\mu }$.
Proton decay proceed through operators with $d=5$. Dominant decay mode is
$p\to K\nu $ with life time of the order of $10^{28}$ years (compare with
experimental bound $\tau (p\to K\nu)> 10^{32}$years).
From the ratio of electron and muon masses we get $m_s/m_d=200$
which contradicts phenomenological value $m_s/m_d=20\div 30$.

Both these dissapointing results follow from
one source: Yukawa interactions of
quark-lepton (super)multiplets with Higgs fields in the minimal SU(5).
Beyond minimal model Yukawa interactions
are less constrained. In our approach
pattern of quark and lepton masses and CKM matrix is explained by the higher
dimension operators through which first two fermion generations get their
masses. Now predictions for proton life time differ drastically from that
of minimal SUSY SU(5) GUT. Since only third generation fermions interacts with
$5$ and $\bar 5$ Higgs fields in the same way as in minimal model,
operators with $d=5$ involve these heavy particles which can not participate
in proton decay. Bare third generation particles get admixtures from
first  two generations which are small. This smallness guarantee
smallness of the deviation of CKM matrix from unity. In this way proton
decay is also suppressed. For scalar quark masses $m_{\tilde{q}}=500$~GeV
we obtain: $\tau (p\to K\nu )\sim
10^{32}$years
which is 4 orders of magnitude better than in minimal model. It is
interesting to note that $p\to K\nu $ decay proceed due to large mass of
top quark which manifest itself in noticeable mass difference between
$\tilde{t}$ and $\tilde{u}$, $\tilde{c}$. To suppress proton decay further
two possibilities exists. First, straightforward one uses heavier squarks. In
this way  $p\to K\mu $ mode dominates over additionally suppressed
($\sim m_t^2/m_{\tilde{q}}^2$) $p\to K\nu $ mode and for
$m_{\tilde{q}}= 1.2$~TeV and
$m_{\tilde{W}}=100$~GeV we  get
$\tau (p\to K\mu )= 10^{32}$years.
Second possibility is intimately connected with desirable ratio
$m_{\mu }/m_s\approx \pm 3$ at GUT scale. There are two possibilities
to get this ratio in our model: to form the 45-plet from the product
of Higgs fields $24\times \bar 5$ (famous Georgi-Jarlskog construction)
or to compose 45-plet and 5-plet in a special way. In the first case
contribution to $d=5$ operator is of the order of $\lambda_s$ while
in the second case it is suppressed. It equals zero for $m_{\mu }/m_s=3$
and is less then $0.1\lambda_s$ for experimentally acceptable choice
~$m_{\mu }/m_s=2.6-3.4$. In this way even for
$m_{\tilde{q}}=600$~GeV and $m_{\tilde{W}}=100$~GeV  we
get $\tau_p=10^{32}$ years.

Search for proton decay at Superkamiokande detector should define future
fate of the suggested scenario.

\vspace {3mm}

{\bf {Acknowledgement}}

We thank Francesco Vissani who has participated in the earlier stage
of this work.
M.V. is grateful to G. Fiorentini and L. Piemontese for hospitality
in Ferrara, where this work was done. The
investigations of M.V. were supported by INTAS grant 93-3316,
INTAS-RFBR grant 95-0567, RFBR grants 96-02-18010 and 96-15-96578.

Z.B. and M.V. are grateful to the organizers of the extended workshop
on Highlights in Astroparticle Physics during which final version of
the paper was prepared and reported.

\newpage

\newpage
\begin{figure}
\begin{center}

             %  FIG. 1a

\begin{picture}(500,100)(10,70)

\put (196,0){\line(1,0){44}}
\put (280,0){\line(1,0){44}}
\put (196,80){\line(1,0){44}}
\put (280,80){\line(1,0){44}}
\put (300,0){\line(0,1){20}}
\put (300,0){\line(0,1){20}}
\put (220,60){\line(0,1){20}}
\put (220,0){\line(0,1){20}}
\put (300,60){\line(0,1){20}}

\put (160,0){\vector(1,0){36}}
\put (160,80){\vector(1,0){36}}
\put (260,0){\line(1,0){20}}
\put (260,80){\line(1,0){20}}
\put (260,0){\vector(-1,0){20}}
\put (260,80){\vector(-1,0){20}}
\put (360,0){\vector(-1,0){36}}
\put (360,80){\vector(-1,0){36}}
\put (220,40){\vector(0,1){20}}
\put (300,40){\vector(0,1){20}}
\put (220,40){\vector(0,-1){20}}
\put (300,40){\vector(0,-1){20}}
\put (140,40){(a)}

\put (165,4){$\nu_{\mu }(u)$}
\put (165,84){$d(s)$}
\put (223.2,60){$\tilde {T}$}
\put (223.2,20){$\tilde {\bar T}$}
\put (303.2,40){$\tilde {W}$}
\put (253,4){$\tilde {s}(\tilde {\mu })$}
\put (253,84){$\tilde {c}$}
\put (344,84){$s(d)$}
\put (344,4){$u(\nu_{\mu })$}

%\put (255,-2.4){$\times $}
%\put (255,77.6){$\times $}
\put (214.5,40){$\times $}
\put (294.5,40){$\times $}

\end{picture}
\vspace{2cm}
\begin{picture}(500,100)(10,70)

                                   %  FIG. 1b
\put (36,24){\line(3,2){24}}
\put (36,56){\line(3,-2){24}}
\put (60,40){\line(1,0){60}}
\put (120,40){\line(3,2){15}}
\put (120,40){\line(3,-2){15}}
\put (180,0){\line(1,0){24}}
\put (180,80){\line(1,0){24}}
\put (180,0){\line(0,1){80}}
\put (180,0){\line(-3,2){15}}
\put (180,80){\line(-3,-2){15}}

\put (0,0){\vector(3,2){36}}
\put (0,80){\vector(3,-2){36}}
\put (240,0){\vector(-1,0){36}}
\put (240,80){\vector(-1,0){36}}
\put (150,60){\vector(-3,-2){15}}
\put (150,60){\line(3,2){15}}
\put (150,20){\vector(-3,2){15}}
\put (150,20){\line(3,-2){15}}
\put (-5,40){(b)}

\put (184,40){$\tilde {W}$}
\put (12,0){$s$}
\put (12,76){$\nu _{\mu }$}
\put (224,4){$u$}
\put (224,84){$d$}
\put (89.6,42.4){$\bar T$}

\put (174,40){$\times $}
%\put (146,16){$\times $}
%\put (143,55){$\times $}
\put (146,63){$\tilde {c}$}
\put (146,5){$\tilde {d}$}
%\put (157,-5){$\tilde {d}^{*}$}
%\put (163,75){$\tilde {c}^{*}$}

                               %  FIG. 1c

\put (296,24){\line(3,2){24}}
\put (296,56){\line(3,-2){24}}
\put (320,40){\line(1,0){60}}
\put (380,40){\line(3,2){15}}
\put (380,40){\line(3,-2){15}}
\put (440,0){\line(1,0){24}}
\put (440,80){\line(1,0){24}}
\put (440,0){\line(0,1){80}}
\put (440,0){\line(-3,2){15}}
\put (440,80){\line(-3,-2){15}}

\put (260,0){\vector(3,2){36}}
\put (260,80){\vector(3,-2){36}}
\put (500,0){\vector(-1,0){36}}
\put (500,80){\vector(-1,0){36}}
\put (410,60){\vector(-3,-2){15}}
\put (410,60){\line(3,2){15}}
\put (410,20){\vector(-3,2){15}}
\put (410,20){\line(3,-2){15}}
\put (505,40){(c)}

\put (444,40){$\tilde {W}$}
\put (272,0){$s$}
\put (272,76){${\nu }_{\mu }$}
\put (484,4){$u$}
\put (484,84){$d$}
\put (349.6,42.4){$\bar T$}

\put (434,40){$\times $}
%\put (266,16){$\times $}
%\put (263,55){$\times $}
\put (385,55){$\tilde {c}$}
\put (387,17){$\tilde {s}$}
%\put (277,-2){$\tilde {s}^*$}
%\put (283,75){$\tilde {c}^*$}

\end{picture}

\vspace{5cm}
\caption
{This picture represents  Feynman diagrams which contribute into the
leading in
minimal SUSY SU(5) decay mode $p\to K^+{\nu }_{\mu }$.
Sum of the contribution
of Fig. 1b and Fig. 1c
equals zero.}
\end{center}
\end{figure}
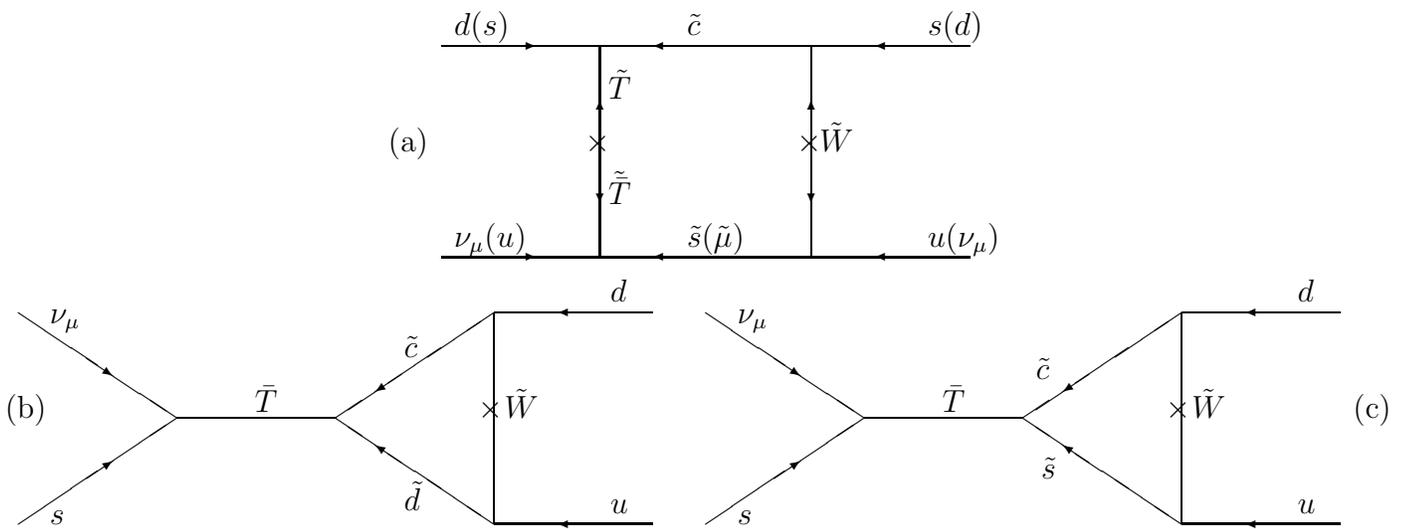

\begin{figure}
\begin{center}

                                    %  FIG. 2a
\begin{picture}(500,100)(10,70)

\put (36,24){\line(3,2){24}}
\put (36,56){\line(3,-2){24}}
\put (60,40){\line(1,0){60}}
\put (120,40){\line(3,2){15}}
\put (120,40){\line(3,-2){15}}
\put (180,0){\line(1,0){24}}
\put (180,80){\line(1,0){24}}
\put (180,0){\line(0,1){80}}
\put (180,0){\line(-3,2){15}}
\put (180,80){\line(-3,-2){15}}

\put (0,0){\vector(3,2){36}}
\put (0,80){\vector(3,-2){36}}
\put (240,0){\vector(-1,0){36}}
\put (240,80){\vector(-1,0){36}}
\put (150,60){\vector(-3,-2){15}}
\put (150,60){\line(3,2){15}}
\put (150,20){\vector(-3,2){15}}
\put (150,20){\line(3,-2){15}}

\put (-5,40){(a)}
\put (184,40){$\tilde {W}$}
\put (12,0){$d^{\gamma }$}
\put (12,76){${\nu }^{\rho }$}
\put (224,4){$u^{\delta }$}
\put (224,84){$d^{\alpha }$}
\put (89.6,42.4){$\bar T$}

\put (174,40){$\times $}
%\put (146,16){$\times $}
%\put (143,55){$\times $}
\put (146,63){$\tilde {u}^{\beta }$}
\put (146,5){$\tilde {d}^{\sigma }$}
%\put (157,-5){$\tilde {d}^{*\sigma }$}
%\put (158,75){$\tilde {u}^{*\beta }$}

                                    %  FIG. 2b

\put (296,0){\line(1,0){44}}
\put (380,0){\line(1,0){44}}
\put (296,80){\line(1,0){44}}
\put (380,80){\line(1,0){44}}
\put (400,0){\line(0,1){20}}
\put (400,0){\line(0,1){20}}
\put (320,60){\line(0,1){20}}
\put (320,0){\line(0,1){20}}
\put (400,60){\line(0,1){20}}

\put (260,0){\vector(1,0){36}}
\put (260,80){\vector(1,0){36}}
\put (360,0){\line(1,0){20}}
\put (360,80){\line(1,0){20}}
\put (360,0){\vector(-1,0){20}}
\put (360,80){\vector(-1,0){20}}
\put (460,0){\vector(-1,0){36}}
\put (460,80){\vector(-1,0){36}}
\put (320,40){\vector(0,1){20}}
\put (400,40){\vector(0,1){20}}
\put (320,40){\vector(0,-1){20}}
\put (400,40){\vector(0,-1){20}}

\put (265,4){${\nu }^{\rho } $}
\put (265,84){$d^{\gamma }$}
\put (323.2,60){$\tilde {T}$}
\put (323.2,20){$\tilde {\bar T}$}
\put (403.2,40){$\tilde {W}$}
\put (360,4){$\tilde {d}^{\sigma }$}
%\put (380,4){$\tilde {d}^{*\sigma }$}
\put (360,84){$\tilde {u}^{\beta }$}
%\put (380,84){$\tilde {u}^{*\beta }$}
\put (444,84){$d^{\alpha }$}
\put (444,4){$u^{\delta }$}
\put (490,40){(b)}

%\put (355,-2.4){$\times $}
%\put (355,77.6){$\times $}
\put (314.5,40){$\times $}
\put (394.5,40){$\times $}

\end{picture}

\vspace{1.6cm}
                                    %  FIG. 2c
\begin{picture}(500,100)(10,70)

\put (36,24){\line(3,2){24}}
\put (36,56){\line(3,-2){24}}
\put (60,40){\line(1,0){60}}
\put (120,40){\line(3,2){15}}
\put (120,40){\line(3,-2){15}}
\put (180,0){\line(1,0){24}}
\put (180,80){\line(1,0){24}}
\put (180,0){\line(0,1){80}}
\put (180,0){\line(-3,2){15}}
\put (180,80){\line(-3,-2){15}}

\put (0,0){\vector(3,2){36}}
\put (0,80){\vector(3,-2){36}}
\put (240,0){\vector(-1,0){36}}
\put (240,80){\vector(-1,0){36}}
\put (150,60){\vector(-3,-2){15}}
\put (150,60){\line(3,2){15}}
\put (150,20){\vector(-3,2){15}}
\put (150,20){\line(3,-2){15}}

\put (-5,40){(c)}
\put (184,40){$\tilde {W}$}
\put (12,0){$d^{\alpha }$}
\put (12,76){$u^{\delta }$}
\put (224,4){${\nu }^{\rho }$}
\put (224,84){$d^{\gamma }$}
\put (89.6,42.4){$ T$}

\put (174,40){$\times $}
%\put (146,16){$\times $}
%\put (143,55){$\times $}
\put (146,63){$\tilde {u}^{\beta }$}
\put (146,5){$\tilde {e}^{\rho }$}
%\put (157,-5){$\tilde {e}^{*\rho }$}
%\put (163,75){$\tilde {u}^{*\beta }$}

                                    %  FIG. 2d

\put (296,0){\line(1,0){44}}
\put (380,0){\line(1,0){44}}
\put (296,80){\line(1,0){44}}
\put (380,80){\line(1,0){44}}
\put (400,0){\line(0,1){20}}
\put (400,0){\line(0,1){20}}
\put (320,60){\line(0,1){20}}
\put (320,0){\line(0,1){20}}
\put (400,60){\line(0,1){20}}

\put (260,0){\vector(1,0){36}}
\put (260,80){\vector(1,0){36}}
\put (360,0){\line(1,0){20}}
\put (360,80){\line(1,0){20}}
\put (360,0){\vector(-1,0){20}}
\put (360,80){\vector(-1,0){20}}
\put (460,0){\vector(-1,0){36}}
\put (460,80){\vector(-1,0){36}}
\put (320,40){\vector(0,1){20}}
\put (400,40){\vector(0,1){20}}
\put (320,40){\vector(0,-1){20}}
\put (400,40){\vector(0,-1){20}}

\put (265,4){$u^{\delta }$}
\put (265,84){$d^{\alpha }$}
\put (323.2,60){$\tilde {T}$}
\put (323.2,20){$\tilde {\bar T}$}
\put (403.2,40){$\tilde {W}$}
\put (360,4){$\tilde {e}^{\rho }$}
%\put (380,4){$\tilde {e}^{*\rho }$}
\put (360,84){$\tilde {u}^{\beta }$}
%\put (380,84){$\tilde {u}^{*\beta }$}
\put (444,84){$d^{\gamma }$}
\put (444,4){${\nu }^{\rho }$}
\put (490,40){(d)}

%\put (355,-2.4){$\times $}
%\put (355,77.6){$\times $}
\put (314.5,40){$\times $}
\put (394.5,40){$\times $}

\end{picture}

\vspace{1.6cm}
                                    %  FIG. 2e

\begin{picture}(500,100)(10,70)

\put (36,24){\line(3,2){24}}
\put (36,56){\line(3,-2){24}}
\put (60,40){\line(1,0){60}}
\put (120,40){\line(3,2){15}}
\put (120,40){\line(3,-2){15}}
\put (180,0){\line(1,0){24}}
\put (180,80){\line(1,0){24}}
\put (180,0){\line(0,1){80}}
\put (180,0){\line(-3,2){15}}
\put (180,80){\line(-3,-2){15}}

\put (0,0){\vector(3,2){36}}
\put (0,80){\vector(3,-2){36}}
\put (240,0){\vector(-1,0){36}}
\put (240,80){\vector(-1,0){36}}
\put (150,60){\vector(-3,-2){15}}
\put (150,60){\line(3,2){15}}
\put (150,20){\vector(-3,2){15}}
\put (150,20){\line(3,-2){15}}

\put (-5,40){(e)}
\put (184,40){$\tilde {W}$}
\put (12,0){$d^{\beta }$}
\put (12,76){$u^{\alpha }$}
\put (224,4){$e^{\rho }$}
\put (224,84){$ u^{\delta }$}
\put (89.6,42.4){$T$}

\put (174,40){$\times $}
%\put (146,16){$\times $}
%\put (143,55){$\times $}
\put (146,63){$\tilde {d}^{\sigma }$}
\put (146,5){$\tilde {\nu }^{\rho }$}
%\put (157,-5){$\tilde {\nu }^{*\rho }$}
%\put (158,75){$\tilde {d}^{*\sigma }$}

                                    %  FIG. 2f

\put (296,0){\line(1,0){44}}
\put (380,0){\line(1,0){44}}
\put (296,80){\line(1,0){44}}
\put (380,80){\line(1,0){44}}
\put (400,0){\line(0,1){20}}
\put (400,0){\line(0,1){20}}
\put (320,60){\line(0,1){20}}
\put (320,0){\line(0,1){20}}
\put (400,60){\line(0,1){20}}

\put (260,0){\vector(1,0){36}}
\put (260,80){\vector(1,0){36}}
\put (360,0){\line(1,0){20}}
\put (360,80){\line(1,0){20}}
\put (360,0){\vector(-1,0){20}}
\put (360,80){\vector(-1,0){20}}
\put (460,0){\vector(-1,0){36}}
\put (460,80){\vector(-1,0){36}}
\put (320,40){\vector(0,1){20}}
\put (400,40){\vector(0,1){20}}
\put (320,40){\vector(0,-1){20}}
\put (400,40){\vector(0,-1){20}}

\put (265,4){$d^{\beta }$}
\put (265,84){$u^{\alpha }$}
\put (323.2,60){$\tilde { T}$}
\put (323.2,20){$\tilde {\bar T}$}
\put (403.2,40){$\tilde {W}$}
\put (360,4){$\tilde {\nu }^{\rho }$}
%\put (380,4){$\tilde {\nu }^{*\rho}$}
\put (360,84){$\tilde {d}^{\gamma }$}
%\put (380,84){$\tilde {d}^{*\gamma }$}
\put (444,84){$u^{\delta }$}
\put (444,4){$e^{\rho }$}
\put (490,40){(f)}

%\put (355,-2.4){$\times $}
%\put (355,77.6){$\times $}
\put (314.5,40){$\times $}
\put (394.5,40){$\times $}

\end{picture}

\vspace{1.6cm}
                                    %  FIG. 2g
\begin{picture}(500,100)(10,70)

\put (36,24){\line(3,2){24}}
\put (36,56){\line(3,-2){24}}
\put (60,40){\line(1,0){60}}
\put (120,40){\line(3,2){15}}
\put (120,40){\line(3,-2){15}}
\put (180,0){\line(1,0){24}}
\put (180,80){\line(1,0){24}}
\put (180,0){\line(0,1){80}}
\put (180,0){\line(-3,2){15}}
\put (180,80){\line(-3,-2){15}}

\put (0,0){\vector(3,2){36}}
\put (0,80){\vector(3,-2){36}}
\put (240,0){\vector(-1,0){36}}
\put (240,80){\vector(-1,0){36}}
\put (150,60){\vector(-3,-2){15}}
\put (150,60){\line(3,2){15}}
\put (150,20){\vector(-3,2){15}}
\put (150,20){\line(3,-2){15}}

\put (-5,40){(g)}
\put (184,40){$\tilde {W}$}
\put (12,0){$e^{\rho }$}
\put (12,76){$u^{\delta }$}
\put (224,4){$u^{\alpha }$}
\put (224,84){$d^{\beta }$}
\put (89.6,42.4){$\bar T$}

\put (174,40){$\times $}
%\put (146,16){$\times $}
%\put (143,55){$\times $}
\put (146,63){$\tilde {u}^{\gamma }$}
\put (146,5){$\tilde {d}^{\sigma }$}
%\put (157,-5){$\tilde {d}^{*\sigma }$}
%\put (158,75){$\tilde {u}^{*\gamma }$}
%\vskip0.7truecm

                                    %  FIG. 2h

\put (296,0){\line(1,0){44}}
\put (380,0){\line(1,0){44}}
\put (296,80){\line(1,0){44}}
\put (380,80){\line(1,0){44}}
\put (400,0){\line(0,1){20}}
\put (400,0){\line(0,1){20}}
\put (320,60){\line(0,1){20}}
\put (320,0){\line(0,1){20}}
\put (400,60){\line(0,1){20}}

\put (260,0){\vector(1,0){36}}
\put (260,80){\vector(1,0){36}}
\put (360,0){\line(1,0){20}}
\put (360,80){\line(1,0){20}}
\put (360,0){\vector(-1,0){20}}
\put (360,80){\vector(-1,0){20}}
\put (460,0){\vector(-1,0){36}}
\put (460,80){\vector(-1,0){36}}
\put (320,40){\vector(0,1){20}}
\put (400,40){\vector(0,1){20}}
\put (320,40){\vector(0,-1){20}}
\put (400,40){\vector(0,-1){20}}

\put (265,4){$e^{\rho }$}
\put (265,84){$u^{\delta }$}
\put (323.2,60){$\tilde { T}$}
\put (323.2,20){$\tilde {\bar T}$}
\put (403.2,40){$\tilde {W}$}
\put (360,4){$\tilde {u}^{\omega }$}
%\put (380,4){$\tilde {u}^{*\omega }$}
\put (360,84){$\tilde {d}^{\gamma }$}
%\put (380,84){$\tilde {d}^{*\gamma }$}
\put (444,84){$u^{\alpha }$}
\put (444,4){$d^{\beta }$}
\put (490,40){(h)}

%\put (355,-2.4){$\times $}
%\put (355,77.6){$\times $}
\put (314.5,40){$\times $}
\put (394.5,40){$\times $}

\end{picture}

\vspace{3cm}
\caption
{All possible diagrams which describe proton decay through
wino dressing. }

\end{center}
\end{figure}
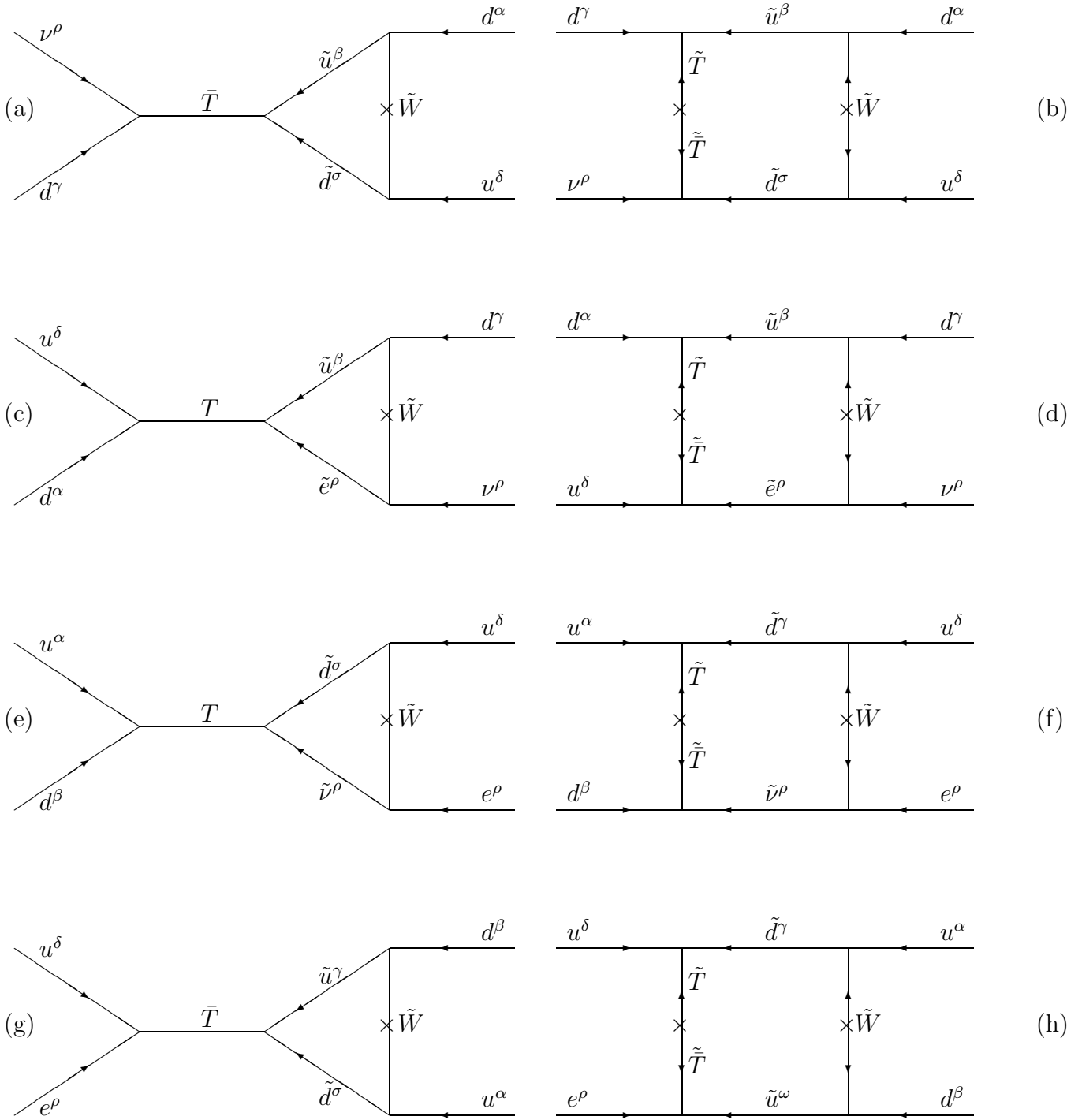

\end{document}